\input{aipcheck}

\documentclass[
    ,final            
  ]
  {aipproc}

\layoutstyle{8x11single}

\begin{document}

\title{Gauged Flavor, Supersymmetry and Grand Unification\footnote{Invited talk presented at the GUT 2012 workshop, Kyoto, Japan, 2012}}

\classification{12.15.Ff}
\keywords      {Gauged flavor, vector-like quarks, grand unification}

\author{Rabindra N. Mohapatra}{
  address={Maryland Center for Fundamental Physics and
Department of Physics, University of Maryland, College Park, MD
20742, USA}}

\begin{abstract}
 I review a recent work  on gauged flavor with left-right symmetry, where all masses and all Yukawa couplings
 owe their origin to spontaneous flavor symmetry breaking. This is suggested as a precursor to a full understanding of flavor of quarks and leptons.
 An essential ingredient of this approach is the existence of heavy vector-like fermions, which is the home of flavor, which subsequently 
 gets transmitted to the familiar quarks and leptons via the seesaw mechanism.
I then discuss  implications of extending this idea to include supersymmetry and finally speculate on a possible grand unified model based on
the gauge group $SU(5)_L\times SU(5)_R$  which provides a group theoretic origin for the vector-like fermions.
 \end{abstract}

\maketitle



Understanding the flavor of quarks and leptons is a major unsolved problem of the standard model (SM). It is generally believed that
a full understanding of the flavor symmetry $U(3)^5$ that emerges in the limit of zero Yukawa couplings of SM and how it breaks may hold the key to this problem.
Use of these symmetries  also forms the basis for a  recent surge of interest in the so-called minimal flavor violation hypothesis\cite{chivu}, which states that
the reason why SM provides such a good account of observed flavor violation is that any beyond the standard model (BSM) physics that incorporates new Higgs doublets
as a way to understand flavor, must have all Yukawa couplings transform as $(3,\bar{3}, 1)$, $(3,1, \bar{3})$ under the full quark sector chiral flavor group
$U(3)^3$. If this hypothesis is taken literally, one is left with the choice to imagine that the Yukawa couplings of the SM are
merely vev's of a set of spurion scalar fields of a higher scale theory and  that all Yukawa couplings
are vevs of scalar fields of higher scale dynamics. In such an approach,
 to avoid Goldstone bosons from creating conflict with cosmological observations, one must assume that the flavor symmetry of SM is indeed a gauge symmetry.
 In this article, we explore this gauged flavor approach. The approach is however intrinsically different from the usual MFV models\cite{chivu} in that there are no extra standard model Higgs doublets but rather heavy vector like fermions which carry all the flavor information.

A convenient implementation of gauge flavor within SM has been carried out recently\cite{GRV} where it is assumed that quark  masses may owe their origin to a seesaw mechanism, involving vector like quarks and leptons, similar to the neutrinos. The idea of using seesaw mechanism for quarks was already discussed in the literature\cite{qseesaw}more than two decades ago. The interesting point made in ref. \cite{GRV} is that the same quark-seesaw framework also allows for gauging of the flavor symmetry without any anomalies.  The requirement that there be
vector-like quarks at high scale- preferably in the TeV range also implies that the model can
be probed in collider searches.  It was further pointed out in \cite{GRV} that consistent with current flavor changing neutral current constraints, part of the gauged flavor dynamics as well as parts of the vector-like quark spectrum can be probed  in the colliders.

The work of \cite{GRV} is however phenomenologically incomplete since  the model  did not address the issue of neutrino masses. It is however easy to see how a simple extensions  to the lepton sector can be carried out. It
 requires that there be three right handed neutrinos, which is an interesting consequence of flavor gauging since it makes neutrino mass natural . For earlier examples of models where flavor gauging implies non-zero neutrino masses see for instance\cite{kuchi}. The problem then is that in minimal models of this type, tiny Yukawa couplings are needed to give small masses to the neutrinos and even if we accept that, the lepton mixing angles vanish, making the model unacceptable. Clearly some nontrivial extension is needed.
 
At the conceptual level, one finds that not all fermion mass terms in the theory
are protected by a gauge symmetry and could therefore be
arbitrary (even Planck scale, as e.g. the electron mass in QED). This is different from the SM
where  all fermion masses owe their origin to spontaneous symmetry breaking and their magnitude must  therefore be limited by the gauge symmetry breaking scale.
 To solve both these problems,
  it was proposed in \cite{GMS} that we extend the standard model gauge
group to the left-right symmetric group based on $SU(2)_L\times
SU(2)_R\times U(1)_{B-L}$\cite{GMS}. There are then no free mass parameters in the fermion sector of the theory and
all masses arise out of spontaneous symmetry breaking like in the standard model. It was shown that indeed the
lightest vector-like fermions as well the flavor gauge bosons do
indeed remain in the TeV to sub-TeV range even in this extension. An additional advantage
of the left-right version is that it provides a solution to the
strong CP problem\cite{babu1} without the need for an axion. In this talk I review this
work and comment on possible extension of this idea to supersymmetry and grand unification.
The latter may answer the question as to where  the vectorlike quarks
with the particular quantum numbers come from ? It has been known
for some time that if one considers a grand unified extension of
the standard model based on $SU(5)_L\times SU(5)_R$\cite{GUT55},
then the vector like $SU(2)_{L,R}$ singlet quarks and leptons are
automatically part of the fermion spectrum.  This is particularly suited
to accomodate the left-right version of the model \cite{GMS} rather the GRV version.
 \footnote{This is different from
recent attempts to grand unify such models in \cite{feldman}}

\section{Gauged Flavor with Left-Right Symmetry} \label{sec:model}

In the SM, once the Yukawa couplings are set to zero,
the maximal flavor symmetry group is $SU(3)_{Q_L} \times
SU(3)_{u_R}\times SU(3)_{d_R}\times SU(3)_{\ell_L}\times
SU(3)_{\ell_R}$. If the weak gauge group is extended
to that of the left-right symmetric model, the flavor group becomes
$SU(3)_{Q_L} \times SU(3)_{Q_R}\times SU(3)_{\ell_L}\times
SU(3)_{\ell_R}$ which is more economical and, unlike the SM, also
simultaneously explains neutrino masses.

We will therefore start with the gauge group $G_{LR} \equiv
SU(3)_{c} \times SU(2)_L \times SU(2)_R \times U(1)_{B-L} \times
SU(3)_{Q_L} \times SU(3)_{Q_R}\times SU(3)_{\ell_L}\times
SU(3)_{\ell_R}$, where $SU(3)_{Q_L} \times SU(3)_{Q_R}$
represents the flavor gauge symmetries respectively in
the left- and right-handed quark sector, and $SU(3)_{\ell_L}
\times SU(3)_{\ell_R}$ the corresponding ones for the lepton sector.
The particle content and its transformation properties under
fundamental representations of the group $G_{LR}$ are as in
table I below.
\begin{table}
\small
\begin{tabular}{ |c| c |  c| c |c |c |c |c |c |}
   &$SU(2)_L$ & $SU(2)_R$ & $U(1)_{B-L}$ & $SU(3)_{Q_L}$ &
  $SU(3)_{Q_R}$ & $SU(3)_c$ & $SU(3)_{\ell_L}$ & $SU(3)_{\ell_R}$ \\ 
\hline \hline
  $Q_L$ & 2 & & $\frac{1}{3}$ & 3 &   & 3 & & \\ 
  $Q_R$ & & 2 & $\frac{1}{3}$ &   & 3 & 3 & & \\ 
  $\psi^{u}_{L}$ & & & $\frac{4}{3}$   &   & 3 & 3 & & \\ 
  $\psi^{u}_{R}$ & & & $\frac{4}{3}$   & 3 &   & 3 & & \\ 
  $\psi^{d}_{L}$ & & & $-\frac{2}{3}$  &   & 3 & 3 & & \\ 
  $\psi^{d}_{R}$ & & & $-\frac{2}{3}$  & 3 &   & 3 & & \\ \hline
  $L_L$ & 2 & & $-1$ & & & & 3 &   \\ 
  $L_R$ & & 2 & $-1$ & & & &   & 3 \\ 
  $\psi^{e}_{L}$ & &   & $-2$ & & & &   & 3 \\ 
  $\psi^{e}_{R}$ & &   & $-2$ & & & & 3 &   \\ 
  $\psi^{\nu}_{L}$ & &   & 0  & & & &   & 3 \\ 
  $\psi^{\nu}_{R}$ & &   & 0  & & & & 3 &   \\ \hline
  $\chi_L$ & 2 & & 1 &  &  & \\ 
  $\chi_R$ & & 2 & 1 &  &  & \\ 
  $Y_u$ &  &  &  & $ \bar{3}$ & 3 & \\ 
  $Y_d$ &  &  &  & $\bar{3}$ &  3  & \\ 
  $Y_\ell$ & & & & & & &$\bar{3}$ &  3\\
$Y_\nu$ & & & & & & &$\bar{3}$ &  3\\\hline
\end{tabular} 
\caption{Model content for fermions and Higgs bosons.}
\label{tab:model}
\end{table}
\normalsize
It is easy to verify that this field content
makes $G_{LR}$ completely anomaly free. In fact
the full anomaly free gauge group also contains chiral color
$SU(3)_{c,L}\times SU(3)_{cR}$ and if this symmetry is broken at the TeV scale,
it can give rise to near TeV mass axi-gluons\cite{gf} which is a class of particle being searched for at the LHC.
 Our detailed phenomenological considerations
below do not depend on whether axigluons exist or not.

The Yukawa couplings of the model are given buy:
\begin{equation}
\label{eq_Lquarks}
{\cal{L}}_{\rm q} = {\cal{L}}_{\rm q}^{\rm kin} - V(Y_u, Y_d,
\chi_L, \chi_R) + \lambda_u (\bar{Q}_L \tilde \chi_L \psi^{u}_{R}
+ \bar{Q}_R \tilde \chi_R \psi^{u}_{L}) + \lambda_d (\bar{Q}_L
\chi_L \psi^{d}_{R} +
\bar{Q}_R \chi_R \psi^{d}_{L})  \\
 + \lambda_u' \bar{\psi}^{u}_{L} Y_u \psi^{u}_{R} + \lambda_d'
\bar{\psi}^{d}_{L} Y_d \psi^{d}_{R} + {\rm h.c.}~,
\end{equation}
We note at this point that, since under parity $Q_L\leftrightarrow Q_R$
and $\psi^{u}_{L} \leftrightarrow \psi^{u}_{R}$ (and similarly for
$\psi^d_{L,R}$), parity symmetry requires
$Y_{u,d} \leftrightarrow Y^\dagger_{u,d}$ and the $\lambda_{u,d}$
as well as $\lambda'_{u,d}$ couplings to be real.%

Concerning the breaking of the gauge groups, the flavor gauge
group $SU(3)_{Q_L} \times SU(3)_{Q_R}$ is broken spontaneously
by the vevs of $Y_u$ and $Y_d$ while the group $SU(2)_L \times SU(2)_R$
by the vevs of the Higgs doublets, $\chi_{L,R}$, as already mentioned.
In particular, we adopt the following vev normalization
$ <\chi_L> = \left(
\begin{array}{c}
 0 \\ v_L
\end{array}
\right)~,~~~~~
<\chi_R> =
\left(
\begin{array}{c}
 0 \\ v_R
\end{array}
\right)~,
$
while diagonal $Y$ vevs will be denoted henceforth as $<\hat Y_{u,d}>$. It is the 
$<Y_{u,d}>$'s which are responsible for fermion masses and mixings. Thus all flavor
originated from flavor breaking.

\subsection{Fermion masses}

From eq.\ (\ref{eq_Lquarks}) one can read off the up-type fermion mass Lagrangian to be
${\cal L}_m ~=~ \bar { U}_L M_u  U_{R}$, with $U ^T= (u, \psi^{u})$, each of the
$u$ and $\psi^{u}$ fields carrying a generation index. The mass matrix reads
\begin{eqnarray}
\label{eq_Mud}
M_u =
 \left(
\begin{array}{cc}
0
& \lambda_u v_L I_{3 \times 3}   \vspace{3mm}
\\
 \lambda_u v_R I_{3 \times 3}
& \lambda_u' < \hat Y_u>
\end{array}
\right)~,~~~~~
M_d =
 \left(
\begin{array}{cc}
0
& \lambda_d v_L I_{3 \times 3}   \vspace{3mm}
\\
 \lambda_d v_R I_{3 \times 3}
& \lambda_d' < \hat Y_d>
\end{array}
\right)~.
\end{eqnarray}
For simplicity, let us work in the limit that the parameters $\lambda_u v_L$
and $\lambda_u v_R$ are much smaller than any of the
$\lambda'_u < \hat Y_u >_i$. Without loss of generality we can assume that $<Y_d>$ is diagonal and $<Y_u>=V_{CKM}\hat{Y}_uV^\dagger_{CKM}$>
(With the subscript $i$ in
$< \hat Y_{u(d)} >_i$ we shall henceforth label the diagonal
entries of the flavon vev matrices.). We can then do a leading order diagonalization
of quark fields and get quark masses by first changing to a basis where $\psi^\prime_u=V^\dagger_{CKM}\psi_u$ and $u^\prime=V^\dagger_{CKM}u$.
 Then, to leading order in an
expansion in the parameters, the masses of the up and down quartks can be written as:
$m_u=\frac{\lambda_u v_{L(R)}}{\lambda_u\hat{Y_u}}$.
From this, the CKM matrices that govern the weak mixings in the light quark sector
 are inherited from
off-diagonalities in the flavon vevs $<Y_{u,d}>$.
In the absence of exact parity, we will  have a flavon vev pattern of the form
$
< Y_u> ~=~ V^\dagger_R < \hat Y_u> V_L~,
~~~~~< Y_d> ~=~ < \hat Y_d>~,
$
with $V_{L,R}$ unitary. We summarize the salient points of the above discussion below:

\begin{itemize}

\item[(i)] In the limit of $v_R \ll < \hat Y_{u,d} >_i$ the elements
of the diagonal $< \hat Y_{u,d} >$ matrices follow an inverted hierarchy
with respect to the quark masses \cite{GRV,qseesaw}.

\item[(ii)] For a given value of $v_R$ and of the $\lambda^{(\prime)}$ couplings,
 or the corresponding exact expressions allow to fix the $< \hat Y_{u,d} >$ entries.
Since the $Y_{u,d}$ vevs set also the mass scale for the flavor gauge bosons, the inverted hierarchy
mentioned in item {(i)} implies a similar hierarchy in new flavor changing
neutral current effects: the lighter the generations, the more suppressed
the effects \cite{GRV}. This is arguably one of the most attractive features
of the model and has been quantitatively analyzed in ref.\cite{GMS}. We summarize the results below.

\item[(iii)] The mass matrices $M_{u,d}$ in the above discussion are
hermitian, leading to arg det[$M_{u,d}$]=0, and implying that the strong
CP parameter at the tree level vanishes. The one loop calculation
for a more general case of this type was carried out in Ref.\ \cite{babu1}.
Using this result, we conclude that the model solves the strong CP
problem without the axion.

\end{itemize}
\subsection{Flavor gauge boson masses and phenomenology}

The masses of the $SU(3)_{Q_L} \times SU(3)_{Q_R}$ gauge bosons ${G_i}_{L,R}$ $(i=1,...,8)$ are
obtained from the kinetic terms of $Y_u$ and $Y_d$ in the Lagrangian,
$Tr\left( | D^{\mu} Y_{u,d} |^2 \right)$ 
In this subsections, we will discuss the various observables that are expected
to provide a constraint (or else the possibility of a signal) for the model. Since in some cases
-- starting from the model spectrum -- the model predictions vary in a wide range, we found it
useful to explore these predictions with a flat scan of the model parameters in ref.\cite{GMS} and the 
results are summarized below. This is done for both the cases with and without TeV scale parity symmetry.
In the other scenario where parity is not a good symmetry at the TeV scale, all the left vs. right couplings
can be chosen as different from each other. Concerning the $SU(2)_{L,R}$ couplings, there vare examples of scenarios where
$g_R/g_L \neq 1$ for a UV complete theory which conserves parity.
In ref.\cite{GMS}, we limited ourselves to the  choice $g_R = 0.7 \cdot g_L$, which can be achieved in such models.

The flavor gauge bosons $G^{\mu a}_{L,R}$ couple to the currents $J^{\mu a}_{H L,R} \equiv
g_{H} \overline Q_{L,R} \gamma^\mu \frac{\lambda^a}{2} Q_{L,R}$. Similarly as in Ref.\cite{GRV},
these interactions give rise to new, tree-level, contributions to the 4-fermion operators
\begin{eqnarray}
\label{eq_Qi}
&&Q_1^{q_j q_i} ~=~ (\bar q_i^\alpha \gamma^\mu_L q_j^\alpha)(\bar q_i^\beta \gamma_{\mu,L} q_j^\beta)~, \\
&&\tilde Q_1^{q_j q_i} ~=~ Q_1^{q_j q_i}|_{L \to R}~, \\
&&Q_5^{q_j q_i} ~=~ (\bar q_i^\alpha P_L q_j^\beta)(\bar q_i^\beta P_R q_j^\alpha)~,
\end{eqnarray}
with Latin and Greek indices on the quark fields denoting flavor and respectively color, and
where $P_{L,R} \equiv (1 \mp \gamma_5)/2$. In the quark mass eigenstates basis, the Wilson
coefficients of the above operators read
\begin{eqnarray}
\label{eq_Ci}
C_1^{q_j q_i} &=& - \frac{g_H^2}{8} (M^2_V)^{-1}_{a,b} (V_L^q \lambda^a V_L^{q \dagger})_{ij}
(V_L^q \lambda^b V_L^{q \dagger})_{ij}~, \\\nonumber
\tilde C_1^{q_j q_i} &=& - \frac{g_H^2}{8} (M^2_V)^{-1}_{8+a,8+b} (V_R^q \lambda^a V_R^{q \dagger})_{ij}
(V_R^q \lambda^b V_R^{q \dagger})_{ij}~,\nonumber \\
C_5^{q_j q_i} &=& \frac{g_H^2}{2} (M^2_V)^{-1}_{a,8+b} (V_L^q \lambda^a V_L^{q \dagger})_{ij}
(V_R^q \lambda^b V_R^{q \dagger})_{ij}~,
\end{eqnarray}
where $q$ can be $u$ or $d$, and a sum over $a$ and $b$ in the range $1,...,8$ is understood.
%
Updated bounds on the Wilson coefficients in eq.\ (\ref{eq_Ci}) have been reported by the UTfit
collaboration \cite{UTfit_DF2} and usefully tabulated in their table 4 for the different
meson-antimeson mixing processes. The contributions, predicted in our model, to the above coefficients
have been explored by the random scan mentioned above.
As previously anticipated, these contributions are well within the existing bounds in the bulk of the
explored parameter space. 

For the exact TeV scale parity, meaning $g_R / g_L = 1$), we find
the lower bounds on the masses of the lightest vectorlike fermion (the top partner) and the lowest
allowed mass for the lightest gauge boson to be 5b TeV and 10 TeV's respectively and for the case of 
no TeV-scale parity (where we assume, as mentioned $g_R / g_L = 0.7$ ), both those values come down to the sub-TeV range
and hence accessible at the LHC. For details see, \cite{GMS}. The current ATLAS and CMS bound on the vector-like top partners are
760 GeV and 475 GeV's respectively.

Similarly for the mixings with vector-like quarks,  the first and second generation quark partners have 
 very small mixings with $u,d,c,s$ quarks whereas for the third generation and the right handed top partner,
the mixings can be of order one. This has several interesting consequences for LHC search of vector-like quarks.

\begin{itemize}

\item   In the pp collision, we couild expect a large cross section for the production of a pair of $\psi_t\bar{\psi}_t$, and each vector like quark through its mixing
will decay to $\psi_t\to t+H$ with $t\to b+W$ and $H\to b\bar{b}$ and similarly for the $\bar\psi_t$ leading to a spectacular 
signature of six b-quarks in the final state. 

\item  One would expect large FCNC effects in $t$ decays e.g. $t\to c+g$ is much enhanced over the SM prediction. With the definition
${\cal L}_{eff}=\kappa \bar{t} \sigma_{\mu\nu} c G^{\mu\nu}$. in SM we expect $\kappa \sim 10^{-5}$ (TeV)$^{-1}$ whereas in our model
we expect  $\kappa \sim 10^{-3}\left(\frac{TeV}{Y_{u,33}}\right)^3$ (TeV)$^{-1}$. The current D0 limit on this is $0.018$ TeV$^{-1}$.

\end{itemize}

It has also been pointed out that the presence of the light flavor gauge bosons allows a reconciliation of the "brewing" $\epsilon_K-\sin 2 \beta$ anomaly\cite{buras}.

The same mechanism can be replicated in the lepton sector and  the neutrinos now have mass. The vector-like fermions in the lepton sector include
three vectorlike charged fermions ($E_{i, L,R}$) and heavy neutral leptons ($N_{i,L,R}$). In the presence of these fermions, the flavor gauge group becomes $SU(3)_{\ell,L}\times 
SU(3)_{\ell, R}$ under which the lepton doublets of the LR model as well as the $E,N$ transform as triplets.  The gauge group is then anomaly free. Like in the quark case, there are flavon fields $Y_{\ell, \nu}$ which carry lepton flavor in their vevs. The mass matrices have similar forms as in the quark case. Without any additional Higgs fields, the neutrino are Dirac fermions. 

\section{Flavor pattern from symmetry breaking}
What distinguishes this approach to flavor from other ones in the literature is that all flavor originates from the vev of the flavon fields 
$Y_{u,d}$. It is therefore necessary to say a few words about this\cite{belen,ilmo}. In an unpublished work, we have looked at the minimization of
the flavon potential in this approach. The flavon potential at the renormalizable level can be written as for the $Y_u$ and $Y_d$ as $V_u+V_d+V_{ud}$

\begin{equation}
V_u = - m^ 2_u T r  Y_ u^\dagger Y_ u  + \lambda _1  T r (Y_ u^\dagger Y_ u) ^ 2 + \lambda_ 2 T r ( Y_ u^\dagger Y_ u Y_ u^\dagger Y_ u  + Det{Y_u}
\end{equation}
\begin{eqnarray}
 V_d=- m^ 2_d T r  Y_ d^\dagger Y_ d  + \lambda _1  T r (Y_ d^\dagger Y_ d) ^ 2 + \lambda_ 2 T r ( Y_ d^\dagger Y_ d Y_ d^\dagger Y_ d  + Det{Y_d}
 \end{eqnarray}
\begin{eqnarray}
 V_{ud}~=~+m^2_{ud}Tr(Y^\dagger_uY_d)+\sum_{i,j,k,l}\lambda_{ijkl}Tr(Y^\dagger_iY_jY^\dagger_kY_i)
  +\sum_{i,j,k,l}\lambda^\prime_{ijkl}Tr(Y^\dagger_iY_j)Tr(Y^\dagger_kY_i)+\epsilon^{ijk}Y_iY_jY_k~+h.c.
 \end{eqnarray}
 where $i,j,k,l$ in the third line go over $u,d$ with the understanding that all $u$ and all $d$ terms are omitted. The minimum of this potential
 corresponds to 
 \begin{equation}
 <Y_{u,d}>~=~\left(\begin{array}{ccc} M_{u,d} & 0 & 0\\ 0 & 0 & 0 \\0 & 0 & 0\end{array}\right)
 \end{equation}
 Thus this gives rise to the leading flavon vev, that corresponding to $Y_{u,d, 11}$. Once we include $d=6$ terms in the potential, it induces a smaller vev in the $22$ entries and  the Det- terms then induce the $33$ entrees. With further higher order terms, we can also induce the off diagonal terms, although to get the hierarchical pattern, we need to do fine tuning. One might contemplate generating the higher dimensional terms from a radiative symmetry breaking scheme.  
 
 \section{Lepton sector} 
 
 We now briefly discuss the lepton sector of the model. Before proceeding to the lepton in the left-right symmetric gauged flavor model, let us discuss the situation in the model of ref.\cite{GRV}. While the ref.\cite{GRV} does not discuss the lepton sector, a possible extension of this model  to include the leptons is straight forward and would be to introduce the leptonic flavor gauge groups, $G_{L,H}\equiv SU(3)_{\ell, H}\times SU(3)_{e_R,H}$. This group becomes anomaly free if in addition to SM leptons, we add two $SU(3)_{\ell, H}$ triplet but SM singlet fermions $\psi_{E_R},\psi_{ N_R}$ and an $\psi_{E_L}$ which is a triplet under $SU(3)_{e_R,H}$ group. We also include the flavor Higgs field $Y_\ell (3,\bar{3})$ under $G_{L,H}$. One can then write down the full gauge invariant  Yukawa interaction for leptons and the mass terms to be:
\begin{eqnarray}
{\cal L}_{\ell}~=~h_E\bar{L}H\psi_{E_R}+~h_\nu\bar{L}\tilde{H}\psi_{N_R}~+~\bar{\psi}_{E_R}Y_\ell\psi_{E_L}
\end{eqnarray}
It is now clear that the charged fermion masses arise in this model from the seesaw mechanism via the vev of the $Y_{\ell}$ field where the neutrino mass is simply $m_\nu=h_\nu<H>$. 

As far as neutrinos go, several points are worth noting:

\begin{itemize}

\item  Anomaly freedom requires  the existence of three right handed neutrinos and hence  massive neutrinos (unlike the standard model where the right handed neutrino has to be added by hand). 

\item In the minimal version of the model,  the neutrino is a Dirac fermion. However to get small masses for them, we need to have $h_\nu\sim 10^{-12}$ and less. While phenomenologically, there is nothing wrong with this, such tiny Yukawa coupling needs some explanation. and is generally considered undesirable.

\item Finally, since $<Y_\ell >$ is the only source for flavor mixing, in the minimal version, by a choice of basis, $<Y_\ell >$ can be diagonalized without affecting any other term
in the Lagrangian (since $h_\nu$ matrix is a unit matrix). As a result, there is no mixing among the neutrinos. Thus the minimal version of the GRV model in the lepton sector is not phenomenologically viable. As we see below, extension of the electroweak sector of the model to make it left-right symmetric, cures this problem.

\item The flavor mixing can be generated by extending the Higgs sector to include a $SU(3)_\ell$ sextet scalar which gives a heavy Majorana mass to the right handed neutrinos and hence the mass to the light neutrinos via the seesaw mechanism\cite{seesaw}. In this case, the $<Y_\ell >$ is inversely proportional to the observed light neutrino mass matrix.

\end{itemize}

An implication of the above model is that since the right handed neutrino $N_R$ and the SM leptonic doublet transform as fundamental representation of the same horizontal group, in a left-handed  neutrino interaction
with another neutrino, one can produce right handed sterile neutrinos. Such interactions will then provide a new drain on the energy on the supernova explosion. Considerations similar to the discussion of right handed neutrinos, imply that the leptonic gauge flavor scale in this model must be at least 20 TeV if the gauge flavor coupling is of the order of the weak gauge coupling\cite{bm}. We will see below that these constraints can be avoided in the left-right symmetric gauge flavor models. 

The lepton sector of the left-right gauged flavor model is specified by the fermion assignment of Table I. Again, the neutrinos are Dirac fermions in the minimal version of the model. Following the same procedure as for the quarks, we see that mass matrix for the Dirac neutrinos has the form:
\begin{eqnarray}
M_\nu~=~\left(\begin{array}{cc}\bar{\nu}_L & \bar{N}_L\end{array}\right)\left(\begin{array}{cc} 0 & \lambda v_L\\ \lambda v_R & <Y_\ell>\end{array}\right)\left(\begin{array}{c} \nu_R\\N_R\end{array}\right)
\end{eqnarray}
The neutrino mass is given by: $m_\nu ~\sim \frac{\lambda^2v_Lv_R}{<Y_{\ell}>}$. The Yukawa coupling can now be in more reasonable range depending on the ratio $\frac{v_R}{<Y_{\ell}>}$. For instance, for $\frac{v_R}{<Y_{\ell}>}\sim 10^{-4}$, we find the largest $\lambda\sim 10^{-4}$. 

As far as the supernova bound on the flavor scale is concerned, due to the heavy mass of the right handed the channel that is open for the GRV model is now blocked. There is now of course constraints on the right handed $W_R$ boson. If parity symmetry is not exact at the weak scale, we can dial down the right handed gauge coupling so that the $W_R$ goes down can easily be in the 5-10 TeV range. This also satisfies the BBN constraints on the new interactions.

\section{Supersymmetric generalization}
The model is easily generalized to accomodate TeV scale supersymmetry. In this case, all fields in the Table I become
superfields and each flavon field and Higgs field e.g. $\chi_{L,R}$ are accompanied by their conjugate fields $\bar{\chi}_{L,R}$ so that
the model remains anomaly free.

An immediate issue with susy gauged flavor with TeV susy breaking is that the soft susy breaking mass terms for the fields 
$\bar{Y}_{u,d}$ and $Y_{u,d}$ are in general different and therefore their vevs
differ by  order TeV mass. This has the consequence that the D-terms of the theory induce large mass differences between different
squark flavors which in turn will lead to large flavor changing neutral current effects.  In our model we can choose gauge mediated
origin for the susy breaking using flavor blind messenger fields so that the $\bar{Y}_{u,d}$ and $Y_{u,d}$ soft masses differ only at the three loop level.
As a result, the induced squark flavor mass difference is of order $\sim \alpha M^2_{SUSY}$ thereby keeping the FCNC effect under control. The detailed
implications of this approach is currently under study.

An interesting point of the SUSY embedding of gauged flavor in our model is that in general it restricts the form of the R-parity violating
terms in the superpotential. Since the leptonic and the quark flavor symmetries are separate, the only R-parity violating term at the renormalizable level
is:\begin{eqnarray}\label{RP}
W_{RPV}~=~\lambda^{''}_R \psi^c_u\psi^c_d\psi^c_d+\lambda^{''}_L\psi_u\psi_d\psi_d+\lambda_L LL\psi^c_e+\lambda_R L^cL^c\psi_e
\end{eqnarray}
Note that both the leptonic and the quark couplings in Eq.\ref{RP} are antisymmetric in the family indices, since they must be gauged flavor invariant.
Focusing on the quark sector, we note that prior to symmetry breaking the interactions in \ref{RP} only connects the heavy quarks. 
Once symmetry is completely broken, the heavy quarks will mix with the light quarks and lead to effective R-P breaking terms
of type $u^cd^cd^c$ in the superpotential involving the mass eigenstate quark super-fields. The strength of these interactions are
 given by $\lambda^{''}_{L,R}\left(\frac{v^3_{L,R}}{<\hat{Y}_u\hat{Y}^2_d>}\right)$.
Note that the dominant R-parity violation comes from the $\lambda_R$ term involving SM singlet fields and secondly for the first two generations this contribution
is  highly suppressed. For example a typical expectation for the coupling $\lambda_{321}\sim 10^{-8}$.  This leads to $\Delta B=2$ neutron-anti-neutron oscillation via
the diagram in Fig.1. 
The expected strength of $N-\bar{N}$ oscillation is $\sim \frac{10^{-22}}{M^5_{susy}}\simeq 10^{-37}$, which puts it beyond the reach of contemplated experiments. This property of the gauged flavor models is very similar to the case of MFV models\cite{smith}.
\begin{figure}
  \includegraphics[scale=1.5]{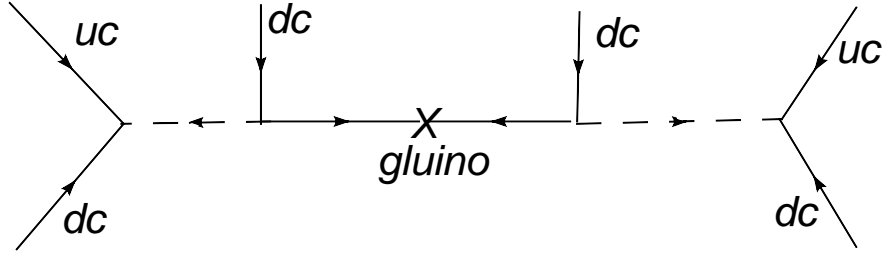}
  \caption{The tree level diagram for neutron-anti-neutron oscillation due to R-parity violating interaction}
\end{figure}


\section{Grand unification possibilities}
The model is based on the left-right symmetric $SU(5)_L\times
SU(5)_R$ gauge group with fermions assigned in a left-right
symmetric manner to the ${\bf 5}\oplus{\bf 10}$ of each group.
They are given below for one of them below and the other follows
by replacing L by R:
\begin{eqnarray}
F_L~=~\left(\begin{array}{c}D^c_1 \\D^c_2\\D^c_3\\e^-
\\\nu\end{array}\right)_L; T_L~=~\left(\begin{array}{ccccc}
0 & U^c_3 &-U^c_2 & u_1 & d_1 \\-U^c_3 & 0 & U^c_1 & u_2 & d_2\\
U^c_2 & -U^c_1 &0&u_3 & d_3\\ -u_1 & -u_2 & -u_3 & 0 & E^{c+}\\
-d_1 & -d_2 & -d_3 & -E^{c+} & 0\end{array}\right)_L
\end{eqnarray}
First point to note is that the quark and charged lepton fields
denoted by $U,D,E$ are the vectorlike fields. They now emerge as
part of the requirement of unification. Second point is that the
maximal anomaly free flavor group that can be gauged is $SU(3)_V$
under which both the left-handed as well as the right handed
fields transform as three dimensional representations.
In order to give mass to the neutrinos, we can add $N_{i, L, R}$ as $SU(3)$ 
triplets and $SU(5)$ singlets.

Turning now to gauge symmetry breaking and fermion masses, we
first note that at the GUT scale, the color SU(3) group is a
chiral group and it breaks down to QCD at some lower scale.

We can envision the symmetry breaking as follows:

(i) {\bf (24,1)}+{\bf (1,24)} to break $SU(5)_L\times SU(5)_R$ down to
$SU(3)_{c,L}\times SU(2)_L\times U(1)_{Y,L}\times
SU(3)_{c,R}\times SU(2)_R\times U(1)_{Y,R}$;

(ii) Use ${\bf ( 5, \bar{5})+(10, \bar{10})}$ (denoted by
$\Sigma_5, \Sigma_{10}$) vevs break $SU(3)_{c,L}\times
SU(3)_{c,R}$ to QCD and also give mass to the vector like quarks
and lepton. It is clear that this will give same mass to all three
flavor partners (U, C, T) of up, charm and top quarks, similarly
for down (D) and charged leptons (E).

As far as the light heavy mixed masses are concerned, since the
mass term is of the form $F_aT_b$ that generates mass terms of the
form $dD^c$ and $F_aF_b$ of the form $uU^c$ (a,b being the flavor
or horizontal quantum numbers), they are flavor nonsinglets. We
therefore use left and right flavon fields $Y_{ab}$ which are
singlets under $SU(5)\times SU(5)$ but sextets under the flavor
group $SU(3)_V$. These masses arise from Yukawa couplings of the
form
\begin{eqnarray}
{\cal L}_Y~=~h_1\bar{F}_LF_R\Sigma_5 +h_2
\bar{T}_LT_R\Sigma_{10}~+~\\ \nonumber (h_3 F_LT_L \bar{H_L}~+~h_4
T_LT_LH_L)\frac{Y^*}{M}~+~L\rightarrow R ~+~h.c
\end{eqnarray}
Here $H_{L,R}$ transform as $(\bf 5,1)\oplus (\bf 1,5)$ under
$SU(5)_L\times SU(5)_R$. After electroweak and right handed
symmetry breaking, the quark-vectorlike quark mass matrices take
the form:

\begin{eqnarray}
{\bf M_{dD}}~=~\left(\begin{array}{cc} 0 & h_3v_Ly_{ab}
\\h_3v_Ry^*_{ab} & h_1<\Sigma_5>\end{array}\right)
\end{eqnarray}
and similarly for the up quarks and the charged leptons. Here
$y_{ab}= \frac{<Y_{ab}>}{M}$. The mass formula for the light down
quarks is then given by
\begin{eqnarray}
{\bf M_d}_{ab}~\simeq \frac{h^2_3v_Lv_R}{h_1<\Sigma_5>}(y^Ty)_{ab}
\end{eqnarray}
As in \cite{GMS}, we will need two flavon multiplets to have
nonzero quark mixings since one flavon vev can always be
diagonalized by an $SU(3)_V$ transformation.

As far as coupling unification is concerned, it is worth pointing out that the GUT scale value of $\sin ^2\theta_W(M_U)=\frac{3}{8(1+\alpha_L/\alpha_R)}$ as noted in
the third paper of \cite{GUT55}. This is to be contrasted with the $\sin ^2\theta_W(M_U)$ values for simple GUT theories e.g. $SU(5)$ or $SO(10)$ 
where it is equal to $\frac{3}{8}$.
This implies that in order to get the weak scale value of $\sin ^2\theta_W(M_U)$, we must have $\alpha_R \gg \alpha_L$ i.e. parity must be broken before right handed gauge symmetry breaks. Typically this requires that the unification scale be much lower than the canonical $10^{15}-10^{16}$ GeV. This raises the question as to whether the model is consistent with current proton life time bounds. The answer to this question is "yes" since the tree level gauge exchange generates typical baryon number violating operators of the form:
\begin{eqnarray}
O_B = \bar{\psi}_{d^c} C^{-1} \gamma^\mu  L \cdot \bar{\psi}_{u^c} \gamma_\mu Q / M^2_U
\end{eqnarray}
In order to generate the proton decay operator, one must use two heavy light mixing factors and with each one being very small, this gives rise to the strength of proton decay
operator which is consistent with current proton decay life time bounds.

\section{summary and outlook} In summary, we have discussed a new approach to the fermion flavor problem where
the introduction of TEV scale vector-like quarks to the standard model have made it possible to gauge
the flavor symmetry. In this framework, all Yukawa couplings arise from flavor symmetry breaking while leaving
some new particles with masses in the TeV range to sub-TeV range. The left-right version of this theory allows a solution
of the strong CP problem without the axion. We also discuss an extension of the model to include supersymmetry and a possible
grand unification is outlined. In the works on the subject to date, the flavon vevs are chosen by hand. Although there is some preliminary work on how to generate them from a complete theory, it will be interesting to  generate the vevs  either from radiative corrections in an UV complete theory or higher dimensional terms so that new insight into the flavor problem can emerge.

\begin{theacknowledgments}
This work is supported 
by National Science Foundation grant No. PHY-0968854. I would like to thank my collaborators on the project, Diego Guadagnoli and Ilmo Sung for many discussions and insights.
\end{theacknowledgments}

\bibliographystyle{aipproc}   

\end{document}